\documentclass[aps,prd,13,onecolumn,preprintnumbers,floatfix,nofootinbib]{revtex4-1}
\pdfoutput=1
\usepackage{graphicx}
\usepackage{bm}
\usepackage{times}
\usepackage{slashed}
\usepackage{color}
\usepackage{slashed}
\usepackage{amsmath}
\usepackage{amsthm}
\usepackage{subfigure}

\newcommand{\be}{\begin{equation}}
\newcommand{\ee}{\end{equation}}
\newcommand{\bea}{\begin{eqnarray}}
\newcommand{\eea}{\end{eqnarray}}

\begin{document}

\title{Muon anomalous magnetic moment in a $SU(4) \otimes U(1)_N$ model without exotic electric charges.}

\author{D. Cogollo$^{a}$}
\email{diegocogollo@df.ufcg.edu.br}

\affiliation{$^a$Departamento de Fisica, Universidade Federal de Campina Grande,
Caixa Postal 10071, 58109-970, Campina Grande, PB, Brazil\\}

\begin{abstract}
We study an electroweak gauge extension of the standard model, so called 3-4-1 model, which does not contain exotic electric charges and it is anomaly free. We discuss phenomenological constraints of the model and compute all the corrections to the muon magnetic moment. Mainly, we discuss different mass regimes and their impact on this correction, deriving for the first time direct limits on the masses of the neutral fermions and charged vector bosons. Interestingly, the model could address the reported muon anomalous magnetic moment excess, however it would demands a rather low scale of symmetry breaking, far below the current electroweak constraints on the model. Thus, if this excess is confirmed in the foreseeable future by the g-2 experiment at FERMILAB, this 3-4-1 model can be decisively ruled out since the model cannot reproduce a sizeable and positive contribution to the muon anomalous magnetic moment consistent with current electroweak limits.
\end{abstract}


\maketitle

\section{Introduction}
\label{intro}

The unified description of the electromagnetic and weak interactions by a single theory, so called Standard Model (SM), certainly is one of the major achievements in this century. The model proposed by Glashow, Salam, and Weinberg in the middle sixties, has been extensively tested during the last decades with tremendous success. The discovery of neutral weak interactions and the production of intermediate vector bosons with the predicted properties increased our confidence in the model. Besides, now the Higgs discovery has anchored, the Standard Model is undoubtedly the best particle physics model we have at our disposal. However, neutrino masses and dark matter are robust experimental and observational indications that the SM must be extended. Furthermore, the excess over the SM prediction on the muon magnetic moment, which is the most precisely measured quantity in particle physics, provides a compelling case for physics beyond the SM. Such issues, have triggered a multitude of extensions of the standard model trying to fully or partially address those matters.\\

In this work we will focus on the muon magnetic moment in the context of an electroweak extension of the SM, called 3-4-1 for short, which is based on the $SU(3)_c SU(4)_L U(1)_X$ gauge symmetry. In general, 3-4-1 models have been proposed to provide an elegant solution to the neutrinos masses, by placing the leptons $\nu,e,\nu^c$ and $e^c$ in the same multiplet of a $SU(4)_L$\cite{Pisano:1994tf}. As in the case of the gauge symmetry $SU(3)_c SU(3)_L U(1)_N$, here the number of fermion families must be an integral multiple of fundamental color which is three, in order to the required anomaly cancellation \cite{Pisano:1991ee}. All these result in an exact family number of three, coinciding with the observation.  Since the third family of quarks transforms under $SU(4)_L$ differently from the first two, this could possibly be the reason to why top quark is so heavy. The $SU(4)_L$ extension can also provide some insights of electric charge quantization observed in the nature \cite{cabarcas2014}. The 3-4-1 model is a natural gauge extension of the so called 3-3-1 models, which are based on the $SU(3)_c \otimes SU(3)_L \otimes U(1)_N$ gauge symmetry. Some of those models provide plausible dark matter candidates in the context of Higgs Portal \cite{331Higgsportal} and $Z^{\prime}$ portal \cite{Mizukoshi:2010ky}, despite the direct dark matter detection controversy \cite{Profumo:2014mpa}, also explaining possibly the Galactic Center excess observed in the Fermi-LAT data \cite{Alves:2014yha}, address the dark radiation non-thermal dark matter production \cite{Kelso:2013nwa}, and even reproducing the mild $H\gamma\gamma$  excess \cite{Alves:2011kc}, among others \cite{othermotiv1,othermotiv2}.\\ 

There are several versions of 3-4-1 models, and each of them inherits the features of their respective 3-3-1 models and therefore such models are in principle  indistinguishable from the 3-3-1 models at low scale. Similarly, if the scale of symmetry breaking of the 3-4-1 model is high enough, at sufficiently low energies 3-4-1 model are equivalent to the SM as well. Albeit, there are remnants in the spectrum that might be important to observables such the muon magnetic moment, which is one of the most precisely measured quantities in particle physics. Differently from other observables the muon magnetic moment might be sensitive to new physics effects at very high energy scales. Several works have been put forth in this direction concerning 3-3-1 models \cite{Kelso:2013zfa}, but there is lack of results in the context of 3-4-1 frameworks \cite{diegomuon}.\\

Our goal here is to assess whether a 3-4-1 model without exotic electric charges and heavy neutral fermions is capable of addressing the excess reported in the muon magnetic moment with respect to the SM prediction, and obtain robust limits in the model in light of the upcoming g-2 experiment at Fermilab, which might reach a $5\sigma$ deviation from the SM.  We will not dwell on unnecessary details. Thus we briefly discuss the key aspects of this model relevant for our reasoning and then present our results.

\section{Model $SU(4)_L  \otimes U(1)_N$}

\subsection{Fermionic Content}

The lepton representations in this model are \cite{palcuD5},
\begin{equation}
\begin{array}{ccccc}
f_{\alpha L}=\left(\begin{array}{c}
\nu_{\alpha}\\
e_{\alpha}\\
N_{\alpha}\\
N_{\alpha}^{\prime}\end{array}\right)_{L}\sim(\mathbf{1,4^{*}},-1/2) &  &  &  & \left(e_{\alpha R}\right)\sim(\mathbf{1,1},-2)\end{array},
\label{Eq.18}
\end{equation}
with $\alpha=1,2,3$. In order to cancel all the quirial anomalies two left handed quark families must transform as 4-plets, and the other one as an anti-4-plet

\begin{equation}
\begin{array}{ccc}
Q_{iL}=\left(\begin{array}{c}
u_{i}\\
d_{i}\\
D_{i}\\
D_{i}^{\prime}\end{array}\right)_{L}\sim(\mathbf{3,4},-1/6) &  & Q_{3L}=\left(\begin{array}{c}
d_{3}\\
u_{3}\\
U\\
U^{\prime}\end{array}\right)_{L}\sim(\mathbf{3},\mathbf{4^{*}},5/6)\end{array}\label{Eq.19}\end{equation}
with $i=1,2$. As for the right handed fields, they transform as:
\begin{equation}
\begin{array}{c}
(d_{3R},(d_{iR}),(D_{iR}),(D_{iR}^{\prime})\sim(\mathbf{3},\mathbf{1},-2/3)\end{array}\label{Eq.20}\end{equation}

\begin{equation}
(u_{3R}),(u_{iR}),(U_{R}),(U_{R}^{\prime})\sim(\mathbf{3},\mathbf{1},4/3)\label{Eq.21}\end{equation}
Similar models have been studied in Refs. \cite{Ponce:2003uu,Doff:2000pc,Ponce:2006vw,Sanchez:2008qv}. In this work we will discuss neither the quark sector nor the corresponding interactions. It is important to point out that the fermions acquire Dirac masses, and in particular for Yukawa couplings of order one, as we will assume, the neutral heavy fermions have mass terms of the type

\begin{equation}
M_{N,N^{\prime}} \simeq V^{\prime}/2
\end{equation}

\subsection{Scalar Sector}
In order to break the symmetry and to give masses to the fermion and gauge bosons in the model, we introduce four scalar multiplets as follows \cite{palcuD5}, 
\begin{eqnarray}\nonumber 
\phi^T_1&=&(\chi^0,\chi^-_1,\chi^-_2,\chi^-_3)=(v^{\prime},0,0,0)\sim[1,4,-3/2], \\ \nonumber
\phi^T_2&=&(\rho^+,\rho^0_1,\rho^0_2,\rho^0_3)=(0,v,0,0)\sim[1,4,1/2], \\ \nonumber
\phi^T_3&=&(\eta^+,\eta^0_1,\eta^0_2,\eta^0_3)=(0,0,V,0)\sim[1,4,1/2], \\ \nonumber
\phi^T_4&=&(\zeta^+,\zeta^0_1,\zeta^0_2,\zeta^0_3)\rangle=(0,0,0,V')\sim[1,4,1/2]. \\ \label{scalars}
\end{eqnarray}

In this setup the pattern of symmetry breaking occurs according to,

\begin{equation}\nonumber
SU(4)_L\otimes  U(1)_X \stackrel{V^\prime}{\longrightarrow} SU(3)_L\otimes U(1)_Z   \stackrel{V}{\longrightarrow}  SU(2)_L\otimes U(1)_Y \stackrel{v+v^\prime}{\longrightarrow}  U(1)_Q.
\end{equation}

Since $M_{W^{\pm}}=\dfrac{g^2}{2}(v^2+v^{\prime 2})$ we have that $\sqrt{v^2+v^{\prime2}} \approx 174Gev$. For now on we will assume the hierarchy $V\sim V^\prime >> v\sim v^{\prime}$.

We point out that more complicated vev assignments could be chosen. Although, our results on the muon magnetic moment will be mostly based on the vector bosons interactions thus not very sensitive to a particular scalar sector.

\subsection{Gauge Sector}

Since we are dealing with the $SU(4)_L \otimes U(1)_N$ gauge group, there are in total 16 gauge bosons ($A_{\mu}^i, {\rm with}\, i=1,2,..16$). In summary the covariant derivative contains the fields,

\begin{equation}
D_{\mu} \supset \frac{1}{\sqrt{2}}\left(\begin{array}{ccccccc}
D_{\mu}^{1} &  & W_{\mu}^{+} &  & K^{+}_{\mu} &  & X_{\mu}^{+}\\
W_{\mu}^{-} &  & D_{\mu}^{2} &  & K_{\mu}^0 &  & X_{\mu}^{0}\\
K^{-}_{\mu} &  & K_{\mu}^{0\ast} &  & D_{\mu}^{3} &  & Y_{\mu}^{0}\\
X_{\mu}^{-} &  & X_{\mu}^{0\ast} &  & Y_{\mu}^{0\ast} &  & D_{\mu}^{4}\end{array}\right),\label{Eq.26}\end{equation}
with $D_{\mu}^{1}=A_{\mu}^{3}/\sqrt{2}+A_{\mu}^{8}/\sqrt{6}+A_{\mu}^{15}/\sqrt{12}$,
$D_{\mu}^{2}=-A_{\mu}^{3}\sqrt{2}+A_{\mu}^{8}/\sqrt{6}+A_{\mu}^{15}/\sqrt{12}$,
$D_{\mu}^{3}=-2A_{\mu}^{8}/\sqrt{6}+A_{\mu}^{15}/\sqrt{12}$, $D_{\mu}^{4}=-3A_{\mu}^{15}/\sqrt{12}$
as diagonal bosons. Notice that in addition to the SM charged current mediated by the W boson, we will have the following new contributions to the muon anomalous magnetic moment

\begin{equation}
{\cal L}^{CC}_l \supset -\frac{g}{\sqrt{2}}\left[\bar{N}^{0}_{L}\gamma^\mu \mu_{L} K_{\mu}^{+} +\bar{N}_{L}^{0 \prime}\gamma^\mu \mu_{L} X_{\mu}^+\right] + H.c,
\label{ccl}
\end{equation}

with $M_{K^{\pm}}^{2}=\dfrac{g^2}{2}(V^2+v^{\prime 2})$ and $M_{X^{\pm}}^{2}=\dfrac{g^2}{2}(V^{\prime 2}+v^{\prime 2})$. As for the neutral gauge bosons it suffices to say that the $4\times4$ mass matrix has a zero eigenvalue corresponding to the photon. For the remainder $3\times3$ matrix we obtain the mass eigenvectors $Z_{\mu}$, $Z_{\mu}^{\prime}$ and $Z_{\mu}^{\prime \prime}$. In the aproximation $V=V^{\prime}$ the field $Z_{\mu}^{\prime \prime}=A_{\mu}^{8}/\sqrt{3}-\sqrt{2/3}A_{\mu}^{15}$ decouples from the other two and adquires a mass $M_{Z_{\mu}^{\prime \prime}}^{2}=g^{2}V^{2}$. The other two eigenvalues are: $M_{Z}^2=\dfrac{M_{W}^{2}}{\cos^2\theta_W}$ and $M_{Z^{\prime}}^2=\dfrac{g^2}{4}V^2$.  As shown in \cite{palcuD5,Ponce:2003uu} $Z_{\mu}^{\prime \prime}$ couples only with exotic fermions, then regarding the muon anomalous magnetic moment only the $Z^{\prime}_{\mu}$ gauge boson matter and contributes with the following neutral current terms \cite{Ponce:2003uu},

\begin{equation}
L \supset -\frac{g}{C_W} \bar{\mu} \gamma^{\mu}[\dfrac{1-3\cos^2\theta_W}{2\sqrt{f_W}}-\dfrac{\cos^2\theta_W}{2\sqrt{f_W}}\gamma^5]\mu Z^{\prime}_{\mu} 
\end{equation}with
\begin{equation}
f_W=3\cos^2\theta_W-1
\end{equation}

\section{Collider and Flavor Bounds}

Bounds stemming from the Z-pole limit the mass of the $Z^{\prime}$ gauge boson. The current measurements exclude $Z^{\prime}$ masses below 2 TeV \cite{Ponce:2003uu}. On the other hand, flavor changing neutral current (fcnc) processes that might be applicable to this model would exclude masses below 11 TeV \cite{Sanchez:2008qv}. The latter is still sensitive to the parametrization scheme used in the quark sector and also to the choose of what family of quarks transforms as an anti-4-plet. Thus we use the former as reference. This limit of 2 TeV on the  $Z^{\prime}_{\mu}$ mass can be translated into a lower limit of 6.2 TeV on the scale of symmetry of the model. Current bounds on $W^{\prime}$ bosons, which are competitive to the $Z^{\prime}$ one, are not directly applicable to our model since the charged gauge bosons $K^{+},X^{+}$ do not interact identically to the SM $W^{\prime}$.\\ 

We have thus far presented the relevant interactions for the muon magnetic moment, we give in the next section a pedagogic introduction to the muon magnetic moment and present our main findings.

\section{Muon Magnetic Moment}

The Dirac equation predicts a muon magnetic moment 

\begin{equation}
\overrightarrow{M}_{\mu} = g_{\mu} \left( \frac{e}{2m_{\mu}} \right) \overrightarrow{S}
\end{equation}

with gyromagnetic ratio $g_{\mu}=2$. However, quantum loop effects lead to a small calculable deviation from $g_{\mu}=2$, the anomalous magnetic moment, parametrized by $a_{\mu} = (g_{\mu}-2)/2$. The SM prediction for the $a_{\mu}$ is generally divided into three parts: electromagnetic (QED), electroweak (EW) and hadronic contributions \cite{PDG}. The QED part, which is by far the dominant contribution in the SM, includes all photonic and leptonic ($e,\mu,\tau$) contributions and has been computed up to four loops and estimated at the 5 loops\cite{5loops,Hanneke:2008tm}. The EW contribution comprises $W^{\pm},Z$ and Higgs bosons, and has been calculated up to three loops. The hadronic contributions are the most uncertain though and can not be calculated by first principles.\\

The current difference $\Delta a_{\mu}=a_{\mu}^{exp} - a_{\mu}^{SM}=295 \pm 81 \times 10^{-11}$ yields a $3.6\sigma$ discrepancy, providing the hint of physics beyond the SM. Nevertheless the large theoretical uncertainties can blur the significance of this discrepancy. The two main theoretical uncertainties are the hadronic vacuum polarization and the hadronic contribution to the light-by-light scattering graph. Improvements in the theoretical side along with the projected experimental sensitivity for the $g-2$ experiment at Fermilab possibly reach

\begin{equation}
\Delta a_{\mu} ({\rm Fermilab} -SM) = (295 \pm 34) \times 10^{-11},
\label{deltaa}
\end{equation}

enhancing the signal up to $5\sigma$ \cite{fermilabproposal}. This discrepancy in Eq.\ref{deltaa} will be referred as future sensitivity for the muon magnetic moment. That being said we computed all corrections to the muon magnetic moment stemming from our model using the public code in Ref.\cite{Queiroz:2014zfa}. In this model those contributions arise from the presence of charged vector bosons $(K^{\pm},X^{\pm})$ in the loop with the neutral fermions $(N, N^{\prime})$ taking the place of the SM neutrinos, and another from the neutral vector boson $(Z^{\prime})$.  The corrections coming from charged and neutral scalars are suppressed because their coupling with the muon are proportional to the muon mass. Hence, they are henceforth neglected.\\

\begin{figure}[!t]
\centering
\includegraphics[scale=0.8]{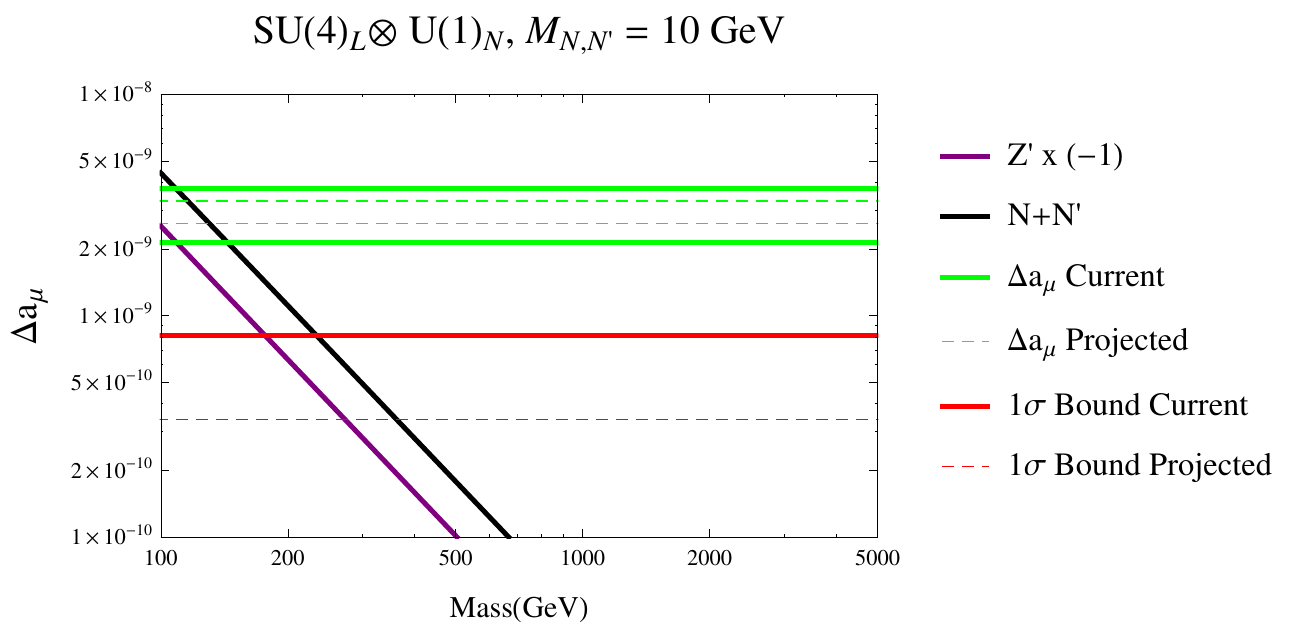}
\caption{Individual contributions to the muon magnetic moment as function of the particles $Z^{\prime}$ and $K^{+},X^{+}$ masses, for heavy neutral fermion masses of $10$~GeV. It is clear that heavy fermions contribution is the leading one.}
\includegraphics[scale=0.8]{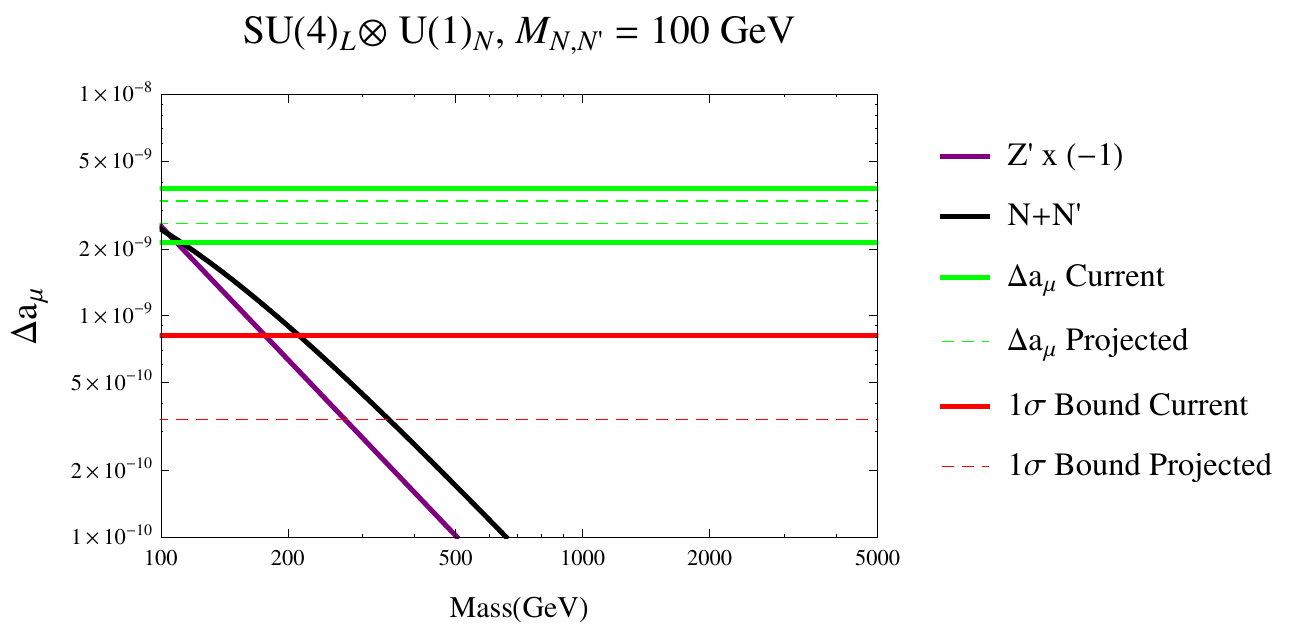}
\caption{Individual contributions to the muon magnetic moment as function of the particles $Z^{\prime}$ and $K^{+},X^{+}$ masses, for heavy neutral fermion masses of $100$~GeV.}
\includegraphics[scale=0.8]{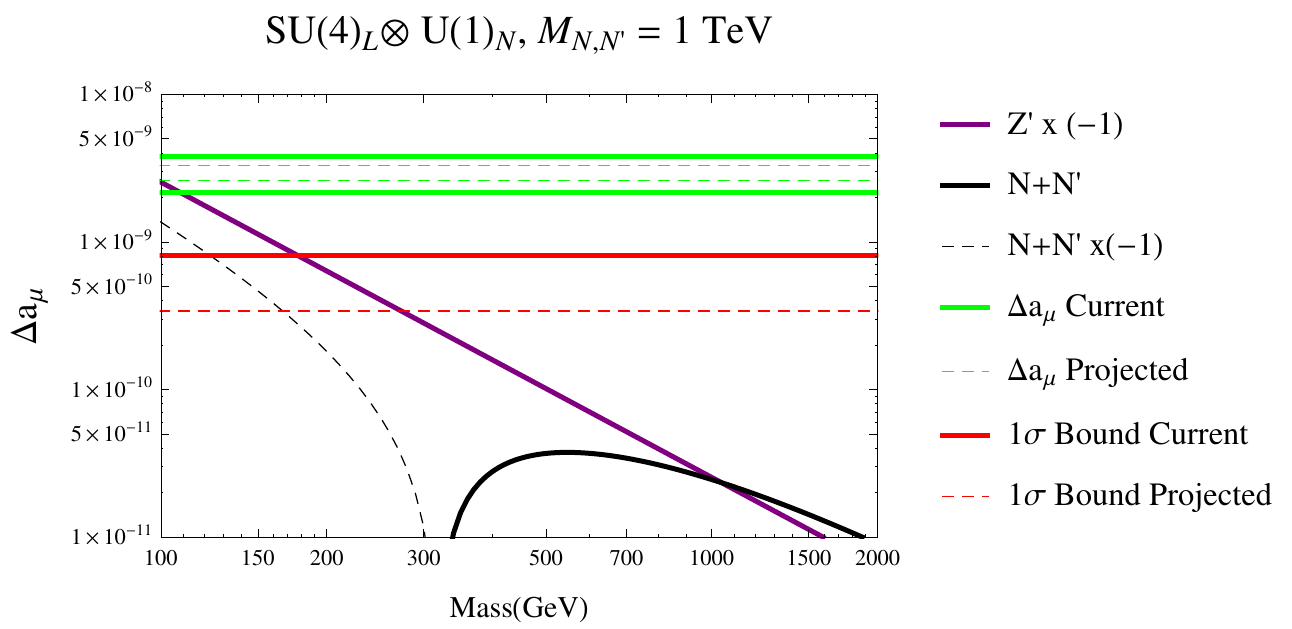}
\label{fig1}
\caption{Individual contributions to the muon magnetic moment as function of the particles $Z^{\prime}$ and $K^{+},X^{+}$ masses, for heavy neutral fermion masses of $1$~TeV. There is a sign change in the correction of the neutral fermions. It is clear that below 300 GeV the $Z^{\prime}$ contribution is more relevant, where for heavier masses the heavy fermions contribution is the leading one.}
\end{figure}

\begin{figure}[!h]
\centering
\includegraphics[scale=0.8]{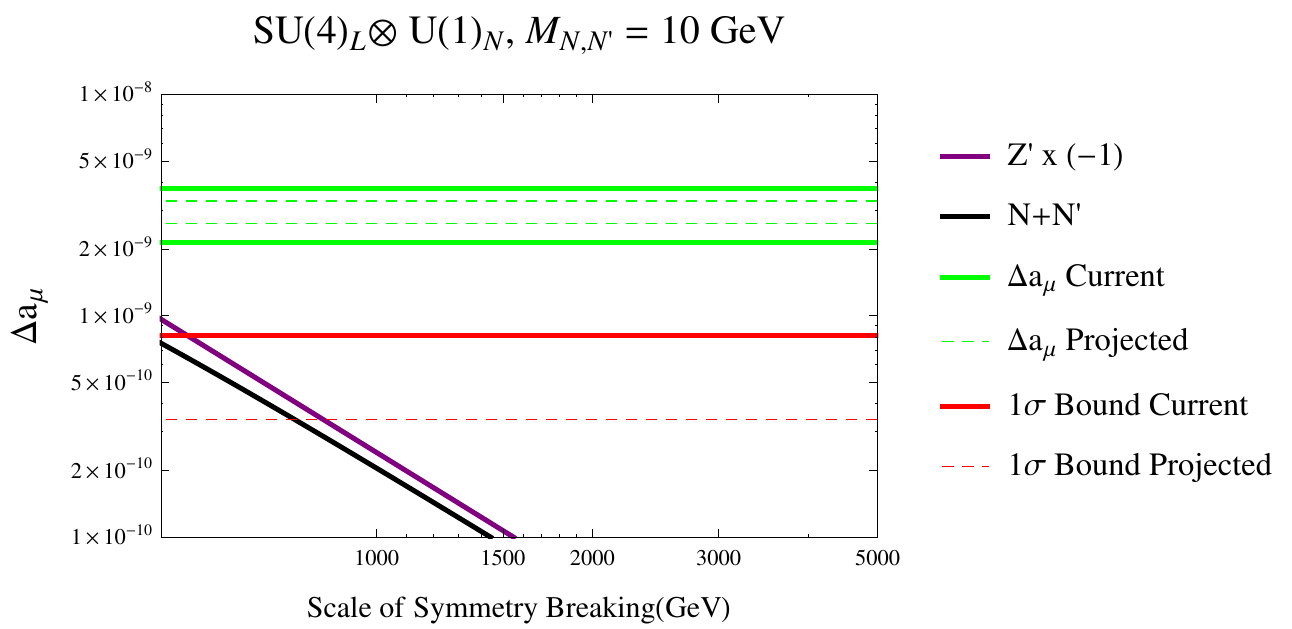}
\caption{Individual contributions to the muon magnetic moment as function of the scale of symmetry breaking, for heavy neutral fermion masses of $10$~GeV. It is clear that heavy fermions contribution is the leading one.}
\includegraphics[scale=0.8]{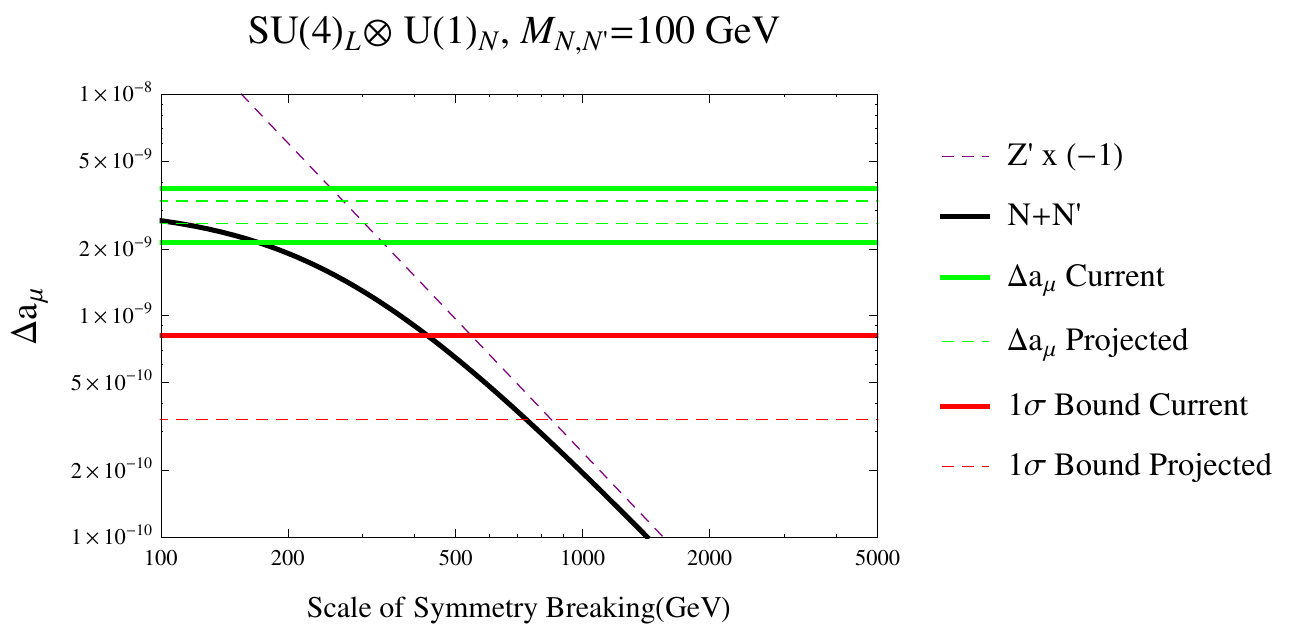}
\caption{Individual contributions to the muon magnetic moment as function of the scale of symmetry breaking, for heavy neutral fermion masses of $100$~GeV. It is clear that heavy fermions contribution is the leading one.}
\includegraphics[scale=0.8]{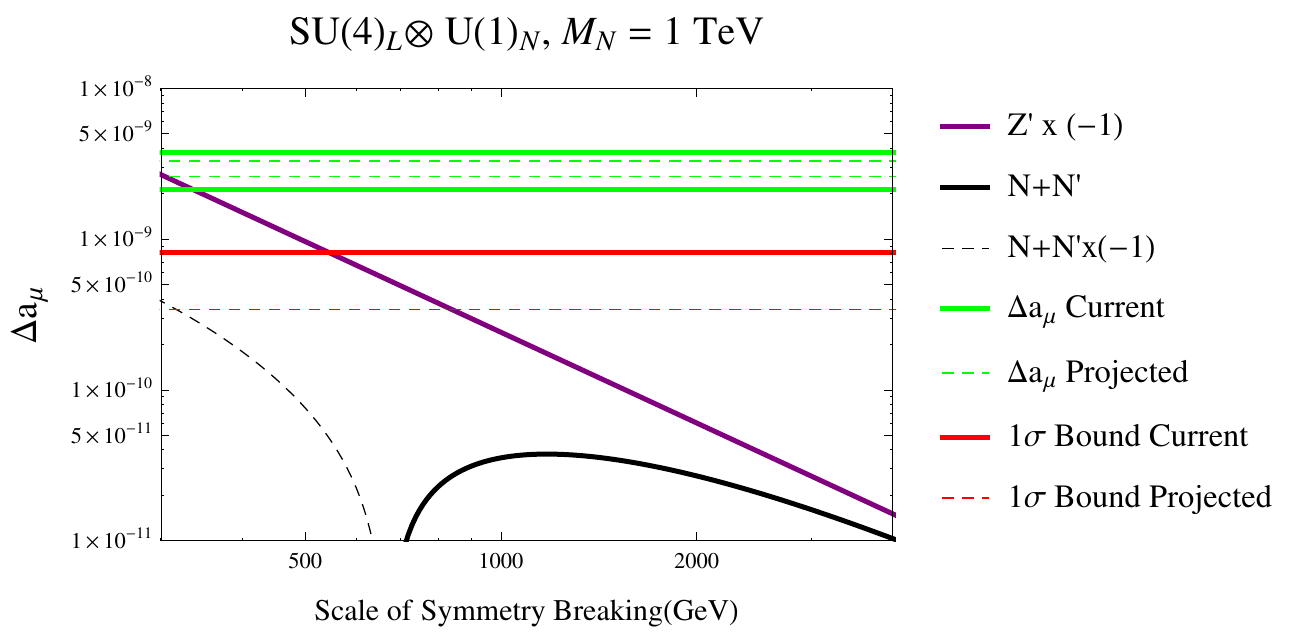}
\caption{Individual contributions to the muon magnetic moment as function of the scale of symmetry breaking, for heavy neutral fermion masses of $1$~TeV. It is clear that heavy fermions contribution is the leading one. The same turning sign feature of Fig.3 happens here.}
\label{fig2}
\end{figure}

In FIG.1-3 we exhibit the individual contributions of each one of those particles as a function of the $Z^{\prime}$ and $K^{\pm},X^{\pm}$ masses for heavy fermion masses of $10$~GeV,$100$~GeV and $1$~TeV respectively. The $Z^{\prime}$ correction is negative, we thus multiplied it by minus one simply to show it. We notice that besides giving a positive correction to the muon magnetic moment, the neutral fermions give rise to the most important contribution for $M_{N,N^{\prime}}=10,100$~GeV, and for charged vector masses of around $150$~GeV the muon magnetic moment excess might be addressed. The masses of the neutral fermions were assumed to be the same. Furthermore, in case the anomaly is otherwise resolved, $1\sigma$ bounds of $\sim 190$~GeV ($\sim 250$~GeV) on the masses of the $Z^{\prime}$  (charged gauge bosons) can be placed. Using projected sensitive these bounds turns to be $\sim 250$~GeV ($\sim 350$~GeV). We emphasize that these bounds are first direct limits on the masses of the charged gauge bosons and neutral fermions in this 3-4-1 model. For $M_{N,N^{\prime}}=1$~TeV though, the $Z^{\prime}$ correction, which is the negative, becomes the leading one for masses below $\sim 1$~TeV. In FIG.3 we see that the neutral fermion contributions become negative for masses below $\sim 300$~GeV. This change in the sign always occurs for sufficiently lower charged vector masses. It did not show in the previous cases because the masses of the neutral fermion were small enough.  Anyhow, we conclude that for $M_{N,N^{\prime}}=1$~TeV the model cannot accommodate the muon magnetic moment discrepancy, and only a projected bound of $160$~GeV and $300$~GeV can be places on the masses of the charged and neutral gauge bosons respectively.\\

In FIGS.4-6 we exhibit the individual contributions of each one of those particles as a function of the scale of symmetry breaking for heavy fermion masses of $10$~GeV, $100$~GeV and $1$~TeV respectively. Since the scale of symmetry breaking is an universal parameter in the model, present in the mass terms of all 3-4-1 particles, it is suitable to depict our findings in terms of this parameter. In FIG.4 the contribution from the neutral and charged gauge bosons are comparable and of opposite sign. Therefore, after adding up those two contributions no relevant limit can be derived on the scale of symmetry breaking. Similarly, for all the remaining cases. The limits on the scale of symmetry breaking derived from the muon magnetic moment are below 1TeV, therefore far below of the aforementioned Z-pole bound. Anyway, we find a very interesting conclusion: if the muon magnetic moment excess is confirmed in the foreseeable future, this model can be decisively ruled out, since there is no way this model can accommodate the muon magnetic moment excess while being consistent with current bounds. Notice that this conclusion is robust, since the main contributions stem from gauge bosons contributions.

\section{Conclusions}

We studied an electroweak gauge extension of the standard model based on the $SU(4)_L \otimes U(1)_N$ gauge symmetry which has heavy neutral fermions in its spectrum, with the purpose of computing the correction to the muon magnetic moment stemming from this model. We find that $Z^{\prime}_{\mu}$ and charged gauge bosons ($K^{\pm},X^{\pm}$) give rise to the most important corrections. The $Z^{\prime}_{\mu}$ corrections are negative due to the relative magnitude of the vector and axial couplings, whereas the ones involving the charged bosons with the heavy neutral fermions can be either positive or negative depending on the masses of the particles involved.  Given the stringent constraint on the scale of symmetry breaking of $\sim 6.2$~TeV, the model decisively cannot accommodate the muon magnetic moment excess in any region of the parameter space. Besides, we derived for the first time in the literature direct upper limits on the masses of the neutral fermions and charged vector bosons that lie in the $100-400$~GeV scale.\\

In summary we conclude that if the muon magnetic moment excess is confirmed at the Fermilab experiment reaching a $5\sigma$ significance, this 3-4-1 model is totally excluded since the model cannot reproduce a sizeable and positive contribution consistent with current electroweak limits.

\acknowledgements

DC is partly supported by the Brazilian National Council
for Scientific and Technological Development (CNPq) Grant
484157/2013-2. We would like to thank Farinaldo Queiroz for useful discussions.

\end{document}